\documentclass[twocolumn,showpacs,aps,prd,amsmath,amssymb,nofootinbib,nobibnotes,floatfix]{revtex4-1}
\usepackage{hyperref}
\usepackage{amsfonts}
\usepackage{amsmath}
\usepackage{mathrsfs}
\usepackage{epsfig}
\usepackage{graphicx}
\usepackage{url}
\usepackage{hyperref}
\usepackage{float}
\usepackage{color}
\usepackage[utf8]{inputenc} 
\usepackage{graphicx}
\usepackage{epsf}
\usepackage{comment}
\usepackage{slashed}
\usepackage{footmisc}
\usepackage{setspace}

\usepackage{hyperref}
\hypersetup{
    pdfnewwindow=true,      
    colorlinks=true,       
    linkcolor=cyan,          
    citecolor=black,        
    filecolor=blue,      
    urlcolor=blue           
}

\begin{document}

\title{How Post-Newtonian Dynamics Shape the Distribution of Stationary Binary Black Hole LISA Sources in nearby Globular Clusters}

\author{Johan Samsing}
\email{Email: jsamsing@gmail.com}
\affiliation{Department of Astrophysical Sciences, Princeton University, Peyton Hall, 4 Ivy Lane, Princeton, NJ 08544, USA.}

\author{Daniel J. D'Orazio}
\email{Email: daniel.dorazio@cfa.harvard.edu}
\affiliation{Department of Astronomy, Harvard University, 60 Garden Street Cambridge, MA 01238, USA}

\begin{abstract}

We derive the observable gravitational wave (GW) peak frequency ($f$) distribution of binary black holes (BBHs) that currently reside inside their globular clusters (GCs),
with and without 2.5 Post-Newtonian (2.5PN) effects included in the dynamical evolution of the BBHs.
Recent Newtonian studies have reported that a notable number of nearby non-merging BBHs, i.e. those BBHs that are expected to undergo further
dynamical interactions before merger, in GCs are likely to be observable by LISA.
However, our 2.5PN calculations show that the distribution of $\log f$ for the non-merging BBH population above $\sim 10^{-3.5}$ Hz scales as $f^{-34/9}$ instead
of the $f^{-2/3}$ scaling found in the Newtonian case. This leads to an approximately two-orders-of-magnitude reduction in the expected number of GW sources at $\sim 10^{-3}$ Hz,
which lead us to conclude that observing nearby BBHs with LISA is not as likely as has been claimed in the recent literature. In fact, our
results suggest that it might be more likely that LISA detects the population of BBHs that will merge before undergoing further interactions. This interestingly suggests
that the BBH merger rate derived from LIGO can be used to forecast the number of nearby LISA sources, as well as providing insight into the fraction
of BBH mergers forming in GCs.

\end{abstract}

\maketitle

\section{Introduction}

Binary black hole (BBH) mergers have recently been observed by the `Laser Interferometer Gravitational-Wave Observatory' (LIGO)
through their emission of gravitational waves \citep{2016PhRvL.116f1102A, 2016PhRvL.116x1103A, 2016PhRvX...6d1015A,
2017PhRvL.118v1101A, 2017PhRvL.119n1101A, 2018arXiv181112907T}, but how and where they formed are still open questions.
Several formation scenarios have been proposed, including active galactic nuclei discs
\citep{2017ApJ...835..165B,  2017MNRAS.464..946S, 2017arXiv170207818M},
isolated field binaries \citep{2012ApJ...759...52D,
2013ApJ...779...72D, 2015ApJ...806..263D, 2016ApJ...819..108B,
2016Natur.534..512B, 2018MNRAS.474.2959G, 2018MNRAS.480.2011G},
single-single GW captures of primordial BHs \citep{2016PhRvL.116t1301B, 2016PhRvD..94h4013C,
2016PhRvL.117f1101S, 2016PhRvD..94h3504C},
dense stellar clusters \citep{2000ApJ...528L..17P,
2010MNRAS.402..371B, 2013MNRAS.435.1358T, 2014MNRAS.440.2714B,
2015PhRvL.115e1101R, 2016PhRvD..93h4029R, 2016ApJ...824L...8R,
2016ApJ...824L...8R, 2016ApJ...824L..12O, 2017MNRAS.464L..36A, 2017MNRAS.469.4665P, 2018MNRAS.478.1844A, 2018arXiv180602351F, 2018MNRAS.479.4652A}, 
galactic nuclei \citep{2009MNRAS.395.2127O, 2015MNRAS.448..754H,
2016ApJ...828...77V, 2016ApJ...831..187A, 2017arXiv170609896H}, and very
massive stellar mergers \citep{Loeb:2016, Woosley:2016, Janiuk+2017,
DOrazioLoeb:2017}. { Among these, two of the most discussed progenitor channels are those resulting from formation in
dynamical environments, such as globular clusters (GCs), and those resulting from formation in
isolation, in the field. However, observationally discerning different formation channels is a non-trivial task, and is therefore the current topic of
many ongoing studies \citep[for a recent review see \textit{e.g.}][]{2018arXiv180605195B}. For example,}
recent work suggests that there are at least two observable parameters that can be used to distinguish these { two classes of} channels. 
The first parameter relates to the angle between the BBH spin vectors that is expected to be random for dynamically formed BBH mergers due to frequent
{ exchanges \citep[\textit{e.g.}][]{2016ApJ...832L...2R}, whereas isolated field BBH mergers are expected to
have somewhat correlated spins \citep[\textit{e.g.}][]{2000ApJ...541..319K, 2008ApJ...682..474B, 2013PhRvD..87j4028G}.} 
Despite its simplicity, this test might fail if, \textit{e.g.}, the field BBH has a third companion \citep[\textit{e.g.}][]{2017ApJ...846L..11L, 2017arXiv171107142A},
or if the individual spin values simply are low.
The second parameter is the BBH orbital eccentricity in the LIGO band, which is expected to be indistinguishable from zero for the field BBH mergers,
but non-zero for a notable fraction of the dynamically formed BBH mergers.
The fraction of such eccentric BBH mergers was only recently derived for GCs in the series of studies by \citep{2006ApJ...640..156G, 2014ApJ...784...71S,
2017ApJ...840L..14S, 2018MNRAS.476.1548S, 2018ApJ...853..140S, 2019MNRAS.482...30S, 2018PhRvD..97j3014S, 2018ApJ...855..124S},
from which it was shown that when { 2.5 Post-Newtonian} (2.5PN) effects \citep[\textit{e.g.}][]{2014LRR....17....2B} are included in the dynamics, $\sim 5\%$ of all GC mergers
are likely to be eccentric in LIGO ({ we refer to eccentric LIGO sources as sources with eccentricity $\gtrsim 0.1$ at $10$ Hz}).
As derived in \cite{2018PhRvD..97j3014S, 2018ApJ...855..124S}, this fraction is $\sim 100$ times greater than
what has been predicted by decades of Newtonian simulations. Eccentric mergers might also form through
single-single GW captures \citep[\textit{e.g.}][]{2009MNRAS.395.2127O, Kocsis:2012ja, 2016PhRvD..94h4013C, 2018ApJ...860....5G},
in environments near super-massive BHs \citep[\textit{e.g.}][]{2018ApJ...853...93R}, and from secular
evolution of Kozai-Lidov triples \citep[\textit{e.g.}][]{2016ApJ...816...65A, 2017ApJ...836...39S, 2018ApJ...864..134R}. In addition,
it has also been argued that relative dense stellar systems might produce second-generation BBH mergers, which are expected to lead
to BHs that have notably higher masses and spins than those formed in
the field \citep[\textit{e.g.}][]{2006ApJ...637..937O, 2016ApJ...824L..12O, 2017PhRvD..95l4046G, 2018PhRvL.120o1101R, 2018ApJ...858L...8C}.
The general BBH mass distribution is likewise expected to differ between different channels, including BBH
mergers forming in GCs \citep{2016ApJ...824L..12O}, primordial BBH mergers \citep[\textit{e.g.}][]{2018ApJ...854...41K}, and single-single GW capture mergers
in galactic nuclei \citep[\textit{e.g.}][]{2018ApJ...860....5G}. Mapping the BBH merger rate as a function of redshift can also help
distinguishing formation channels \citep[\textit{e.g.}][]{2014MNRAS.442.2963K, 2018arXiv180602351F, 2018ApJ...863L..41F}.

Looking towards the future, the `Laser Interferometer Space Antenna' mission \citep[LISA;][]{2017arXiv170200786A}
will be able to provide a wealth of additional insight into how and where BBHs
form \citep[\textit{e.g.}][]{2006ApJ...637..937O, 2016ApJ...830L..18B, 2016PhRvL.116w1102S,
Seto:2016, 2017MNRAS.469..930C, 2017MNRAS.465.4375N, 2018MNRAS.481.5445S, 2018MNRAS.481.4775D}. The LISA mission
operates at much lower GW frequencies than LIGO, and is therefore able to provide insight into the BBH orbital parameters
before { general relativistic effects erase} information about their values near assembly. For example, it was
illustrated by \cite{2018MNRAS.481.5445S, 2018MNRAS.481.4775D} that about $50\%$ of all BBH mergers assembled in GCs
will have a measurable non-zero eccentricity in the LISA band \citep[see also][]{2006ApJ...637..937O} ({ we refer to eccentric LISA sources as
sources with eccentricity $\gtrsim 0.01$ at $10^{-2}$ Hz}), which is orders of magnitude
more than expected for field BBH mergers.

In this paper we study the GW frequency distribution of BBHs that currently are inside their GCs, their possibility for being observed by LISA,
and how these results are affected by the fact that BBHs are likely to merge both during and in-between the 
ongoing dynamical interactions in the GC \citep[\textit{e.g.}][]{2018PhRvL.120o1101R};
processes we here loosely refer to as {\it 2.5PN effects}. Newtonian work on this has been performed
before by \cite{2002CQGra..19.1297B}, in which it was concluded that a few BBHs in our Milky Way should be observable by LISA.
{ Recent work by \cite{2018PhRvL.120s1103K} made similar conclusions using a state-of-the-art
H{\'e}non Monte-Carlo style approach,} but the results were again based on purely
Newtonian dynamics. The importance of including 2.5PN corrections for describing the population of merging BBHs observable by LISA and LIGO was first
described by \cite{2018arXiv180208654S, 2018MNRAS.481.5445S, 2018MNRAS.481.4775D};
however, { at the time of this paper's submission,} no work has discussed the role of 2.5PN corrections for describing the GW frequency distribution of the
currently retained BBH population. We do that here for the first time.

We show that 2.5PN correction effects significantly reduce the probability for observing BBHs near the low frequency end of the LISA band,
exactly where \cite{2018PhRvL.120s1103K} predicts a { nearby observable BBH population}.
{ To shortly review the main results from \cite{2018PhRvL.120s1103K}, it was argued that $\sim 4$, $\sim 8$, and $\sim 80$ BBHs should be
resolvable by LISA with $S/N > 2$ in the Milky-Way, the Andromeda galaxy, and the Virgo-cluster, respectively.}
The work by \cite{2018PhRvL.120s1103K} did
attempt to correct for 2.5PN correction effects by putting an upper limit on the BBH eccentricities, from which it was concluded that the 2.5PN correction are unlikely to play a
role; however, we prove that this estimator insufficiently accounts for 2.5PN induced mergers between dynamical encounters
within the GC. { Using a simple semi-analytical model we derive the
leading order effect from including 2.5PN corrections, from which we conclude that \cite{2018PhRvL.120s1103K} have overestimated the total reported number of
observable sources at least by a factor of a few to an order of magnitude. This difference is found primarily in the the nearby, `Virgo-cluster population' (see above) that
we show are likely to be greatly suppressed due to effects from including 2.5PN corrections}.
Despite the simplicity of our model, we do present the first consistent discussion on this topic,
which indeed suggests that more careful studies on how BBHs evolve and distribute inside GCs must be performed.

The paper is organized as follows. In Section \ref{sec:Modeling Black Hole Dynamics} we describe our approach
to modeling the dynamical evolution of BBHs inside their GCs, and how the inclusion of
2.5PN correction effects affect the dynamics and corresponding observables.
In Section \ref{sec:Results} our main results are presented, which include the
observable distributions of GW peak frequencies for BBHs currently inside their
GCs, and the effects from including 2.5PN corrections in the dynamics. We conclude our study in Section \ref{sec:Conclusions}.

\section{Modeling Black Hole Dynamics}\label{sec:Modeling Black Hole Dynamics}

Our goal is to estimate the probability distribution of GW peak frequencies (see Eq. \eqref{eq:fGW} for a definition) for dynamically assembled BBHs that currently are inside their GCs,
with and without 2.5PN effects. As described in the Introduction,
we use the term `2.5PN effects' to denote that our model allows for in-cluster BBH mergers during and in-between encounters \citep[\textit{e.g.}][]{2018MNRAS.481.5445S}, in contrast
to `Newtonian models' that essentially only allow dynamically assembled BBHs to merge after being ejected
\citep[\textit{e.g.}][]{2018PhRvL.120o1101R, 2018PhRvD..97j3014S, 2018MNRAS.481.5445S}.

For modeling the evolution of BBHs inside GCs, we use our semi-analytical model described in \cite{2018MNRAS.481.5445S}.
In short, in this model we assume that each BBH starts out with a semi-major axis (SMA) equal to its hard binary limit value, $a_{\rm HB}$, which
is given by \citep[\textit{e.g.}][]{Hut:1983js},
\begin{equation}
a_{\rm HB} \approx \frac{3}{2}\frac{Gm}{v_{\rm dis}^{2}},
\end{equation}
where $m$ is the mass of one of the (assumed equal mass) interacting BHs, and $v_{\rm dis}$ is the GC velocity dispersion.
This is a good approximation as the majority of the relevant BBHs are believed to form dynamically through close interactions of $>2$ initially unbound single BHs
in the GC core \citep[\textit{e.g.}][]{Heggie:1975uy, 1976A&A....53..259A, Hut:1983js, 2012NewA...17..272T}; a process which has the highest probability of forming BBHs near $a_{\rm HB}$.
We assume equal masses as both mass segregation and dynamical three-body swappings naturally lead to this limit \citep[\textit{e.g.}][]{2016PhRvD..93h4029R}.
After formation at $a_{\rm HB}$, the BBH undergoes interactions with surrounding single BHs, generally
referred to as `binary-single interactions' \citep[\textit{e.g.}][]{Hut:1983js, 1992ApJ...389..527H}.
Each of these binary-single interactions changes the SMA of the interacting BBH from $a$ to $\delta \times a$,
where the average value of $\delta$ can be shown to equal $7/9$ \citep{2018PhRvD..97j3014S}.
The BBH keeps undergoing binary-single interactions until its SMA $a$ becomes small enough for the three-body recoil to kick it
out of the GC. The critical value for $a$ at which this ejection happens is given by \citep[\textit{e.g.}][]{2018PhRvD..97j3014S},
\begin{equation}
a_{\rm ej} \approx \frac{1}{6} \left(\frac{1}{\delta} - 1\right) \frac{Gm}{v_{\rm esc}^2},
\label{eq:aej}
\end{equation}
where $v_{\rm esc}$ is the escape velocity of the GC.
Depending on the SMA and eccentricity of the ejected BBH, { it will be able to merge within a Hubble time outside of the GC. For example,
for a $30M_{\odot}+30M_{\odot}$ BBH with an initial SMA $a = 0.5$ AU, the initial eccentricity has to be $\gtrsim 0.8$ for the corresponding GW inspiral
time to be less than a Hubble time. Assuming the eccentricity, $e$, distributes according to a so-called thermal distribution P(e) = 2e \citep{Heggie:1975uy},
then one finds that there is about $1-0.8^2 \rightarrow 36\%$ chance for such a BBH to contribute to the observable rate.}
This is the classical way of forming BBH mergers in dense stellar systems \citep[\textit{e.g.}][]{2000ApJ...528L..17P}.
However, when 2.5PN effects are included in the dynamics, the BBH also has a relatively high probability of merging
inside the GC before being dynamically ejected \citep{2018PhRvL.120o1101R, 2018PhRvD..97j3014S, 2018MNRAS.481.5445S}, as summarized below.

When 2.5PN effects are included in the $N$-body dynamics, a BBH undergoing the binary-single hardening sequence described above,
can merge inside the GC in at least two different ways \citep{2018MNRAS.481.5445S}.
The first way is during binary-single interactions (3-body GW merger),
which are generally described by highly chaotic motions \citep{2014ApJ...784...71S}.
As shown in \cite{2018PhRvD..97j3014S, 2018arXiv180208654S}, this population of BBHs forms close to the DECIGO \citep{2011CQGra..28i4011K, 2018arXiv180206977I}
and LIGO bands with { an initial eccentricity close to unity}.
The second path to in-cluster merger is in-between binary-single interactions (2-body GW merger). This can happen if a previous encounter leaves the BBH
with an eccentricity high enough for its GW lifetime \citep[see \textit{e.g.}][]{Peters:1964bc}, denoted by $t_{\rm GW}$, to be smaller than
its binary-single encounter time scale, denoted by $t_{\rm bs}$. About $50\%$ of all BBH mergers in GCs form in this way \citep{2018PhRvL.120o1101R}, and
they { generally spend part of their time in the LISA band with eccentricity $\gtrsim 0.01$ \citep{2018MNRAS.481.5445S, 2018MNRAS.481.4775D},
which interestingly is in the measurable range of LISA}. { Throughout the paper we refer to the in-cluster BBH populations that do and do not merge before their next binary-single
encounter by the `merging population' ($t_{\rm GW} < t_{\rm bs}$) and the `non-merging population' ($t_{\rm GW} > t_{\rm bs}$), respectively.}

\subsection{Numerical Study}\label{sec:Numerical Study}

To derive the GW peak frequency distribution of the retained BBH population, we numerically follow a large set of (uncorrelated) BBHs from
their initial SMA $a_{\rm HB}$ towards their minimum SMA $a_{\rm ej}$, assuming their SMA decreases in each encounter as 
${\delta}^{0}a_{\rm HB}, {\delta}^{1}a_{\rm HB}, {\delta}^{2}a_{\rm HB}, ..., {\delta}^{n}a_{\rm HB}, ..., $ until ${\delta}^{N_{\rm ej}}a_{\rm HB} \approx a_{\rm ej}$,
where $n$ is the $n$'th binary-single interaction, and $\delta = 7/9$.
If 2.5PN effects are not included, all of the BBHs will reach $a_{\rm ej}$, whereas if 2.5PN effects are included
a significant fraction will merge before ejection \citep{2018MNRAS.481.5445S}. Because the 3-body GW
merges contribute to only a few $\%$ of in-cluster mergers, they are not important for this study. Hence, we focus entirely on modeling the effect from 2-body GW mergers.
{ Following the approach from \cite{2018MNRAS.481.5445S}, we do this by calculating at each hardening step $n$ the time between
strong binary-single interactions \citep[\textit{e.g.}][]{2018ApJ...853..140S},
\begin{equation}
t_{\rm bs} \approx  \frac{1}{6 \pi G} \frac{v_{\rm dis}}{n_{\rm s} m a},
\label{eq:tbs}
\end{equation}
and the GW lifetime of the BBH in question,
\begin{equation}
t_{\rm GW} \approx \frac{768}{425}\frac{5c^{5}}{512G^{3}} \frac{a^{4}}{m^{3}}\left(1- e^2 \right)^{7/2},
\label{eq:tGW}
\end{equation}
where $n_{\rm s}$ in Eq. \eqref{eq:tbs} denotes the number density of single BHs, and for Eq. \eqref{eq:tGW} we have used the high eccentricity limit from \citep{Peters:1964bc}.
Furthermore, we assume that the BBH eccentricity, $e$, distributes according to a thermal distribution P(e) = 2e \citep{Heggie:1975uy}.}
We note here that distant encounters (weak binary-single interactions) can also change the eccentricity \citep{1996MNRAS.282.1064H}; however, we do not include this
effect in this paper.
Now, if $t_{\rm GW}<t_{\rm bs}$ at a given step $n$, we stop the interaction series and label the outcome as a 2-body GW merger. If instead the BBH does not merge
before $n = N_{\rm ej}$, the outcome is labeled as an ejected BBH. At each state $n$ and for each outcome, we record the corresponding
BBH orbital parameters. From these we derive the GW peak frequency distribution, as described below.

An eccentric BBH emits GWs with a broad spectrum of frequencies \citep[see \textit{e.g.}][]{2018MNRAS.481.4775D};
however, most of the energy is radiated near the so-called `GW peak frequency' which to leading order is given by \citep[\textit{e.g.}][]{Wen:2003bu},
\begin{equation}
f \approx \frac{1}{\pi} \sqrt{\frac{2Gm}{r_{\rm p}^3}},
\label{eq:fGW}
\end{equation}
where $r_{\rm p} = a(1-e)$ is the pericenter distance of the BBH. Although other frequencies near this value might be observable by LISA
for nearby sources \citep[\textit{e.g.}][]{2002CQGra..19.1297B, 2018MNRAS.481.4775D, 2018PhRvL.120s1103K}, we only discuss implications related to the peak value in this study.

Our numerical results presented in this paper are all based on following $10^{5}$ BBHs from their initial $a_{\rm HB}$
towards $a_{\rm ej}$, for which we assume
that $m = 20M_{\odot}$, $n_{\rm s} = 10^{5}$ pc$^{-3}$, $v_{\rm dis} = 10$ kms$^{-1}$,
and $v_{\rm esc} = 50$ kms$^{-1}$. We note that these values are uncertain; however, they do
result in $\approx 50\%$ of all BBH mergers occurring inside their GC, which is in agreement with
the recent 2.5PN simulations presented in \cite{2018PhRvL.120o1101R}.

We consider what we refer to as the `observable distribution' of BBHs,
by taking into account the probability that a BBH emitting at a given GW frequency $f$ would be observed at a single snapshot in time, i.e. at the time of observation.
This probability is proportional to the time the corresponding BBH spends at that GW frequency $f$, which is equal to $t_{\rm bs}$ for a BBH that will not merge { (non-merg. BBH)},
and { approximately} equal to $t_{\rm GW}$ for a BBH that will merge { (merg. BBH)}. { The last statement follows because a BBH orbit
in the high eccentricity limit evolves with nearly constant peri-center distance, i.e. GW peak frequency, until circularization \citep[\textit{e.g.}][]{2014ApJ...784...71S}.}
Therefore, to produce the observable distribution, we weight each $f$ value derived using our model
by either $t_{\rm bs}$ or $t_{\rm GW}$, as further described in Section \ref{sec:Analytical Scaling Relations}.
Our derived distributions will therefore be directly comparable to the one shown in, \textit{e.g.}, \cite{2018PhRvL.120s1103K}.
We note here that for estimating the actual number of resolvable sources one has to further include the signal-to-noise (S/N), which depends on
the source { including its mass, eccentricity and orientation}, the observational strategy, and instrument \citep[\textit{e.g.}][]{2018MNRAS.481.4775D}. This will not be covered in this paper, but we will comment on it in Section \ref{sec:Conclusions}.

\subsection{Analytical Scaling Relations}\label{sec:Analytical Scaling Relations}

Before presenting our main results, we start here by providing some insight into how the distribution
of $\log f$ can be analytically estimated. 
The derived relations will be compared to our numerical results presented later in Section
\ref{sec:Newtonian Results} and \ref{sec:General Relativistic Results}, and will provide useful
understanding of how 2.5PN effects are expected to affect the observable distributions.

To derive the observable distribution of $\log f$ as a function of $f$, we start by writing out the probability that a BBH will be
observed with a GW peak frequency $>f$ during its dynamical evolution from SMA $a_{\rm ini}$ towards $a_{\rm fin}$,
\begin{equation}
P(>f) \propto \int_{a_{\rm ini}}^{a_{\rm fin}}\frac{p(>f,a)}{a}da,
\label{eq:Pgtf}
\end{equation}
where $p(>f,a)$ denotes the probability that the BBH will be observed with a GW peak frequency $>f$ when its SMA is $=a$.
The above relation originates from summing the probabilities $p(>f,n)$ over the hardening steps from $n=0,..N_{\rm ej}$, which
we have changed to a variation in SMA $a$ using that the change in $a$ in each interaction is $a(1-\delta)$ (see \textit{e.g.} \cite{2018PhRvD..97j3014S}).
Following this notation, the probability distribution of $f$ is therefore given by $P(f) = -dP(>f)/df$, and the distribution in $\log f$ by
$P(\log f) = fP(f)$. This lead us to conclude that,
\begin{equation}
P(\log f) \propto f \frac{d}{df} P(>f) \propto P(>f),
\label{eq:Plogf}
\end{equation}
where the last step is valid only when the solution is a power-law in $f$, which is true in the asymptotic limit we consider.
The probability $p(>f,a)$ can be written as,
\begin{equation}
p(>f,a) \propto \int_{e_{\rm f}}^{1}P(e)W(a,e)de,
\label{eq:pfa}
\end{equation}
where $P(e)$ is the eccentricity distribution, $W(a,e)$ is a weight factor that describes the probability for
observing a BBH with SMA $a$ and eccentricity $e$ at a single snapshot in time, and $e_{\rm f}$
is the eccentricity of a BBH having a GW frequency $f$ and a SMA $a$ given by rearranging Eq. \eqref{eq:fGW},
\begin{equation}
1-e_{\rm f} \propto a^{-1}f^{-2/3}.
\end{equation}
As described in Section \ref{sec:Numerical Study}, the weight factor $W(a,e)$ is to leading order
proportional to the time between binary-single interactions for the non-merging BBHs,
and proportional to the GW lifetime for the merging BBHs, i.e.,
\begin{align}
W(a,e)   & \propto t_{\rm bs},\ 	& \text{(non-merg. BBH)}\label{eq:Wae_tbs}  \\
W(a,e)   & \propto t_{\rm GW},\ 	& \text{(merg. BBH)}\label{eq:Wae_tGW}
\end{align}
With these relations, one can derive the (asymptotic) distribution of $\log f$ with and without 2.5PN effects. This will be illustrated in the sections below.

\section{Results}\label{sec:Results}

In this section we present our main results on the observable distribution of GW peak frequencies for BBHs that
currently are inside their GC. Section \ref{sec:Newtonian Results} presents results from the Newtonian limit, where
Section \ref{sec:General Relativistic Results} shows results from including 2.5PN effects.

\label{sec:Results}
\subsection{Newtonian Results}\label{sec:Newtonian Results}

\begin{figure}
\centering
\includegraphics[width=\columnwidth]{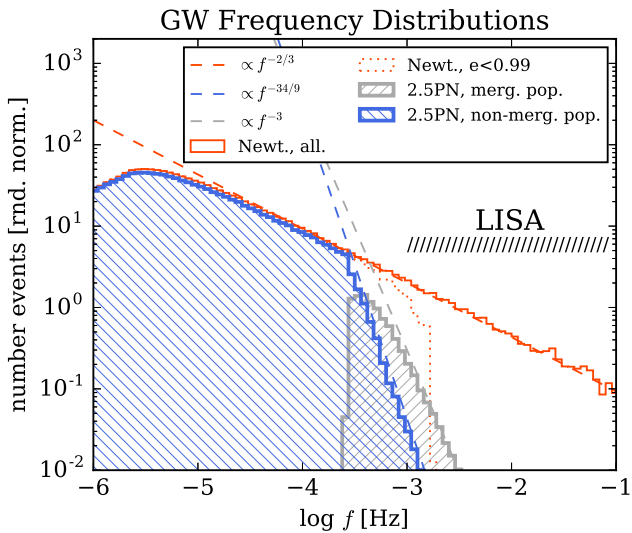}
\caption{
Distribution (y-axis) of GW peak frequencies (x-axis) each weighted by a factor
$W(a,e)$ (see Eq. \eqref{eq:Wae_tbs} and \eqref{eq:Wae_tGW}),
from BBHs that currently are inside their GCs, derived using our model outlined in Section \ref{sec:Numerical Study}. The parameters we
have used are $m = 20M_{\odot}$, $n_{\rm s} = 10^{5}$ pc$^{-3}$, $v_{\rm dis} = 10$ kms$^{-1}$, and $v_{\rm esc} = 50$ kms$^{-1}$.
The figure shows results from dynamically evolving $10^{5}$ BBHs with 2.5PN corrections (blue/grey, `2.5PN') and without 2.5PN corrections (red, `Newt.'), i.e. with
and without taking into account that BBHs can merge in-between their hardening binary-single interactions inside the GC.
The weight factor $W(a,e)$ for a given $f$ is set to be proportional to the time the corresponding BBH spends at that $f$, which is
$ = t_{\rm bs}$ for the non-merging population (`non-merg. pop.'), and $= t_{\rm GW}$ for the merging population (`merg. pop.').
The weight $W(a,e)$ is proportional to the probability for observation, and the figure therefore shows the distribution one would see at a
random snap-shot in time, i.e. at the time of observation (See Section \ref{sec:Modeling Black Hole Dynamics}). { The horizontal hatched region
indicates roughly the window where LISA is most sensitive.}
{\it Red solid:} Distribution without 2.5PN effects (Newt., all.).
{\it Red dotted:} Distribution without 2.5PN effects, and only including BBHs with an eccentricity $<0.99$ (Newt., $e<0.99$).
{\it Blue:} Distribution with 2.5PN effects from BBHs that {do not merge} before their next encounter (2.5PN, non-merg. pop.).
{\it Grey:} Distribution with 2.5PN effects from BBHs that {do merge} before their next encounter (2.5PN, merg. pop.).
The results are discussed in Section \ref{sec:Results}.
}
\label{fig:Pf}
\end{figure}

The observable distribution of $\log f$ derived by the use of our model outlined in Section \ref{sec:Modeling Black Hole Dynamics}
without 2.5PN effects, i.e. not allowing for BBHs to merge inside their GCs, is shown in Figure \ref{fig:Pf} with the {\it red solid} line.
As seen, this Newtonian limit leads to an almost perfect power-law distribution starting from about $10^{-5}$ Hz.
Slightly above $10^{-6}$ Hz the distribution undergoes a clear break, which is at the frequency $f$ corresponding to a BBH with orbital parameters
$a = a_{\rm ej}, e=0$, i.e. at $f(a_{\rm ej},e=0)$. The position of this break therefore scales $\propto v_{\rm esc}^{3}/m$, which follows from Eq. \eqref{eq:aej} and \eqref{eq:fGW},
i.e. its position is not strongly dependent on $m$, but is expected to notably vary with the overall environment through $v_{\rm esc}$.

Our derived distribution seems to be in overall good agreement with the Newtonian results
shown in \cite{2018PhRvL.120s1103K} (Figure 1, black line), \textit{e.g.}, both distributions have a break around $10^{-6}$ Hz and a near power-law decline
towards higher frequencies. The distribution from \cite{2018PhRvL.120s1103K} truncates at about $10^{-2}$ Hz, likely due to low statistics.
This agreement provides some validation of our model, despite its simplicity.

We now derive an analytical scaling solution for $P(\log f)$ without 2.5PN effects, using the framework presented in Section \ref{sec:Analytical Scaling Relations}.
For this, we first use that $W(a,e) \propto t_{\rm bs} \propto 1/a$, which follows from Eq. \eqref{eq:tbs}.
Assuming that $P(e) = 2e$, the probability $p(>f,a)$ is then found from Eq. \eqref{eq:pfa} to
scale $\propto a^{-1}(1-e_{\rm f}^2) \approx a^{-1}(1-e_{\rm f}) \propto a^{-2}f^{-2/3}$, where we here have assumed that $e_{\rm f} \gg 0$.
Inserting this relation into Eq. \eqref{eq:Plogf} results in the solution,
\begin{equation}
P(\log f) \propto f^{-2/3}\ \text{(Newt.)},
\label{eq:PlogfnoGR}
\end{equation}
where `Newt.' refers to our assumption of the Newtonian limit.
This scaling is shown in Figure \ref{fig:Pf} with the {\it red dashed} line. As seen, it perfectly describes the numerically
generated data above $10^{-5}$~Hz, which validates the consistency of our approaches so far. Below we study the effects from including 2.5PN corrections.

\subsection{General Relativistic Results}\label{sec:General Relativistic Results}

We now consider the effects from allowing BBHs to merge through GW emission in-between their binary-single interactions, which is the leading order 2.5PN effect
relevant for this problem.
This inclusion leads to two different observable populations; the population of BBHs that do not merge before their next interaction (non-merging population),
and the population that do merge before undergoing further interactions (merging population).
The corresponding $\log f$ distributions are shown in Figure \ref{fig:Pf} with {\it blue} (2.5PN, non-merg. pop.) and {\it grey} (2.5PN, merg. pop.) lines,
respectively. As seen, the inclusion of 2.5PN effects leads to a sharp break in both the non-merging and merging $\log f$ distributions around $10^{-3.5}$ Hz, which is the
frequency corresponding to $a = a_{\rm ej}$ and $t_{\rm GW} = t_{\rm bs}$, i.e. at $f(a_{\rm ej}, t_{\rm GW}=t_{\rm bs})$.
Using the relations from Section \ref{sec:Modeling Black Hole Dynamics}, one
finds that $f(a_{\rm ej}, t_{\rm GW}=t_{\rm bs}) \propto m^{2/7}v_{\rm esc}^{-12/7}n_{\rm s}^{3/7}$ assuming $v_{\rm dis}v_{\rm esc}^{-1} = $ const.
Therefore, the position of the break seen in Figure \ref{fig:Pf} is expected
to vary only weakly with BH mass, but notably with $v_{\rm esc}$; however, we note that one expects variations from
our chosen values ($m=20M_{\odot}$, $v_{\rm esc} = 50$ kms$^{-1}$) only up to about a factor of two for `standard' GCs and BH masses.

At frequencies higher than the break value, the probability for observing a BBH declines rapidly compared to the Newtonian case (red distribution).
If all the $f$ values are given the same weights as in the Newtonian case, i.e. if we set $W(a,e) = t_{\rm bs}$,
the joint distribution of the two 2.5PN populations (`2.5PN, non-merg. pop.' and `2.5PN, merg. pop.')
is found to be very close to the Newtonian derived distribution.
From this follows that the sharp decline seen for the merging BBHs (`2.5PN, merg. pop.') is directly linked to their observational
weight term $W(a,e) \propto t_{\rm GW} \propto a^{1/2}f^{-7/3}$,
which becomes increasingly smaller than the Newtonian term $W(a,e) \propto t_{\rm bs} \propto 1/a$ for increasing values of the GW peak frequency.
We note here that the cutoff in the merging population (grey distribution) at low frequencies represents
where no value of $e$ could merge the binary within a binary-single interaction time at the given frequency.

The corresponding sharp decline of the non-merging BBH
population (blue distribution) occurs naturally because BBHs with a high GW peak frequency $f$ also have a relative large
probability for merging ($t_{\rm GW} \propto a^{1/2}f^{-7/3}$). Therefore,
the higher the GW peak frequency, the more likely it is for the BBH to merge than to undergo further interactions. If it does undergo further interactions
this means that the time scale $t_{\rm bs} < t_{\rm GW}$ will be relatively short, and thereby its weight term is correspondingly small.
These combined effects give rise to the observed rapid decline in BBH number for increasing $f$.
Below these qualitative descriptions are presented mathematically, where we derive and discuss the two 2.5PN populations.

\subsubsection{The Non-merging BBH Population}\label{sec:The non-merging BBH Population}

For the BBHs that will not merge before their next binary-single encounter (2.5PN, non-merg. pop.), one must have that $P(>f)$ is
non-zero only for values of $a$ greater than the value that fulfills $t_{\rm bs}(a) = t_{\rm GW}(a,f)$. Using
that $t_{\rm GW}(a,f) \propto a^{4}(1-e_{\rm f}^2)^{7/2} \propto a^{1/2}f^{-7/3}$,
and that $t_{\rm bs} \propto 1/a$, one finds this minimum value for $a$ to scale as $\propto f^{14/9}$. Approximating the
non-merging probability by $p(>f,a) \propto a^{-2}f^{-2/3}$, where we have used $W(a,e) \propto t_{\rm bs} \propto 1/a$, 
and evaluating Eq. \eqref{eq:Pgtf} with limits of integration $a_{\rm ini} = a_{\rm HB} \gg a_{\rm fin}$ and $a_{\rm fin} \propto f^{14/9}$, we find the following scaling solution, 
\begin{equation}
P(\log f) \propto f^{-34/9}\ \text{(2.5PN, non-merg. pop.)}.
\label{eq:PlogfwithGRUM}
\end{equation}
As seen in Figure \ref{fig:Pf}, this scaling agrees with the numerically generated data at frequencies above the $10^{-3.5}$~Hz break. At frequencies
below the break, $a_{\rm fin}$ is replaced by $a_{\rm ej}$ and the Newtonian case is recovered as shown above, and as is observed in the figure.

In \cite{2018PhRvL.120s1103K}, an attempt to extract the observable non-merging BBHs from their Newtonian simulations
was done by removing the current BBH population with an eccentricity $>0.99$. However, it is clear from what we have derived so far that
a single cut in eccentricity is insufficient for isolating the non-merging BBHs. For example, for a given BBH with SMA $a$ the corresponding
eccentricity $e_{\rm M}$ above which it will merge is given by solving $t_{\rm bs}(a) = t_{\rm GW}(a,e_{\rm M})$, from which
one finds $1-e_{\rm M} \propto a^{-10/7}$. The value of $a$ changes by one to two orders of magnitude during hardening, which clearly implies the
critical value $e_{\rm M}$ significantly changes as well. To further illustrate this, the {\it red dotted} line in 
Figure \ref{fig:Pf} shows the distribution found from employing the
proposed $0.99$ eccentricity cut. As seen, already at $10^{-3}$~Hz this $e>0.99$ cut (red dotted line) predicts about
two-orders-of-magnitude more non-merging BBHs than our consistent approach (blue solid line).
This calls into question some of the key results from \cite{2018PhRvL.120s1103K},
including that several nearby non-merging BBHs should be observable by LISA, and that 2.5PN effects should not play a major role in this prediction.
{ For example, the Virgo-cluster population that in \cite{2018PhRvL.120s1103K} was predicted to contribute with about $\sim 80$ currently
observable BBH LISA sources comes from up-weighting the small population found with $f > 10^{-3}$~Hz (see Fig. 1 in \cite{2018PhRvL.120s1103K}). However, 
we have in this paper shown, using our 2.5PN dynamical formalism, that the probability for observing these BBHs is very small due to their corresponding short GW lifetime.}
{ Therefore, 2.5PN corrections do indeed play a notable role}, as it clearly leads to a strong break near
$10^{-3.5}$ Hz, above which the observable probability distribution of non-merging BBHs decreases rapidly relative to the Newtonian prediction;
from Eq. \eqref{eq:PlogfnoGR} and \eqref{eq:PlogfwithGRUM} the 2.5PN-derived distribution decreases relative to the Newtonian as
$f^{-34/9}/f^{-2/3} \propto f^{-28/9}$, which is more than three-orders-of-magnitude per decade in $f$.
That said, we do note that our approach only presents an idealized picture due to its simplicity, and observable
outliers and exceptions might exist. However, at the frequencies were LISA is most sensitive, we do still expect
that 2.5PN effects will lead to orders of magnitude differences compared to the Newtonian limit treated in \cite{2018PhRvL.120s1103K}.

\subsubsection{The Merging BBH Population}

For the BBHs that do merge before their next binary-single encounter (2.5PN, merg. pop.), the observational weight factor is proportional to their
GW lifetime, i.e. in this case $W(a,e)\propto t_{\rm GW} \propto a^{4}(1-e^2)^{7/2}$. By including this factor in the integral for $p(>f,a)$
given by Eq. \eqref{eq:pfa}, one finds from integration that $p(>f,a)$ in this case is $\propto f^{-3}$. From plugging this into Eq. \eqref{eq:Pgtf} it now follows that,
\begin{equation}
P(\log f) \propto f^{-3}\ \text{(2.5PN, merg. pop.)}.
\end{equation}
This again agrees with the simulations at frequencies above the $10^{-3.5}$~Hz break. At frequencies below the break, BBHs do not merge
within a binary-single encounter time and the merging population diminishes.

Comparing the non-merging and the merging populations shown in Figure \ref{fig:Pf},
we find that the merging population dominates the potentially observable population at frequencies above the break $10^{-3.5}$ Hz,
right where LISA starts to become sensitive. Although the two distributions are quite similar, this result does hint that the relevant population to
consider for nearby LISA sources might in fact be the population that is on its way to merge. We note that this is the population the authors in
\cite{2018PhRvL.120s1103K} attempted to remove by the $0.99$ eccentricity cut. If the merging population in fact is the most likely to be observed,
one should be able to predict the number of nearby LISA sources by the use of the BBH mergers observable by LIGO. Such a test could further
provide insight into the fraction of BBH mergers assembled in GCs. For more information
on the merging population we refer the reader to \cite{2018MNRAS.481.5445S, 2018MNRAS.481.4775D}. We conclude our study below.

\section{Conclusions}\label{sec:Conclusions}

In this paper we have derived the observable probability distribution of GW peak frequencies
from BBHs that currently are inside their GCs (see Section \ref{sec:Results}).
In particular, we have for the first time characterized the observable effects from taking into account 
the possibility that BBHs can merge in-between their hardening binary-single interactions, which we argue is the leading order effect from including 2.5PN corrections
in this problem (see Section \ref{sec:Modeling Black Hole Dynamics}). We find that 2.5PN corrections strongly suppress the probability for observing
BBHs above $10^{-3.5}$ Hz, which leads us to re-evaluate key results from the recent Newtonian study presented by \cite{2018PhRvL.120s1103K}.

In \cite{2018PhRvL.120s1103K} it was reported that a notable number of
BBHs in GCs are likely to be observable by LISA out to the Virgo cluster\footnote{Note here that BBHs
that are observed drifting through the LISA band and end up in the LIGO band, can be seen out to much larger distances
due to their higher signal-to-noise \citep[\textit{e.g.}][]{2016PhRvL.116w1102S, 2018MNRAS.481.5445S, 2018MNRAS.481.4775D}.}
(similar papers on this include \textit{e.g.} \cite{2002CQGra..19.1297B, 2013LRR....16....4B} and \cite{2017MNRAS.469..930C} for BBHs that are not formed dynamically).
An attempt to correct for 2.5PN effects by removing all current BBHs with
an eccentricity $>0.99$ was also performed, and the corresponding results
led the authors to conclude that 2.5PN effects are likely not to play an important role.
However, we have in this paper illustrated that 2.5PN effects clearly lead to a strong depletion of
observable GW sources above $10^{-3.5}$ Hz, and further
shown that the proposed $0.99$ eccentricity cut is an incorrect estimator.

For the non-merging BBHs, i.e. the population that will undergo further interactions inside their GCs,
we derived that the observable distribution of $\log f$ above $10^{-3.5}$ Hz without 2.5PN effects scales $\propto f^{-2/3}$
and with 2.5PN effects $\propto f^{-34/9}$. This implies that already at $10^{-3}$ Hz the number of observable non-merging BBHs is
about two-orders-of-magnitude smaller when 2.5PN corrections are included compared to what is found in the Newtonian limit, even with the
$0.99$ eccentricity cut proposed by \cite{2018PhRvL.120s1103K}.

Although the non-merging population appeared to be the focus of \cite{2018PhRvL.120s1103K}, our results in fact
hint that it might be more likely that LISA observes the nearby merging BBH population, i.e. the BBHs that will merge before their
next encounter (their GW lifetimes are still relatively long at this stage). This interestingly suggests that the
observed rate of BBH mergers from LIGO can be used to derive the expected number
of nearby BBHs observable by LISA (see similar conclusion for isolated circular BBHs \cite{2016PhRvL.116w1102S} and eccentric
BBHs \citep{2017MNRAS.465.4375N}), which further
can be used to constrain the fraction of BBH mergers forming in GCs.

The importance of including 2.5PN corrections becomes particularly clear when considering the absolute number of nearby BBHs observable by LISA, for which
one has to take into account their associated S/N. The reason is that the S/N for LISA sources generally
increases with $f$ (for $f<10^{-2}$~Hz), which implies that the higher $f$ is the further away the source can be
observed (see \textit{e.g.} \cite{2018MNRAS.481.4775D}). From this follows that the BBHs
in the fully 2.5PN dominated region at $f \gtrsim 10^{-3.0}$~Hz in fact might dominate the observable population in numbers, as their relative high value of $f$
correspondingly gives them a high weight by the many more GCs that are within their larger observable distance. 
For example, as discussed by \cite{2018PhRvL.120s1103K}, there are $\sim 10^{4}$ GCs in the Virgo cluster, compared to $\sim 10^{2}$
in our own Milky Way. Judging from \cite{2018PhRvL.120s1103K}, for a BBH to be observable to the Virgo cluster,
its GW peak frequency must be $\gtrsim10^{-2.5}$~Hz. From our results shown in Figure \ref{fig:Pf}, it is clear
that the effects from 2.5PN corrections fully determine what can be observed near and above that frequency;
the number of nearby BBHs we expect to see with LISA, seems therefore to highly depend on a proper inclusion of 2.5PN effects.

Finally, that the short-lived GW driven BBHs actually seem to play an important role, also brings some concern to the common numerical way of
sampling the BBH distribution using `snapshots'. In \cite{2018PhRvL.120s1103K} the BBH distribution was sampled from snapshots
spaced $10-100$ Myr apart; however, its clear that such an approach will greatly undersample and thereby miss BBHs with relative high GW peak frequencies due to their
associated short GW lifetimes. Therefore, the clear analytical derivations and descriptions we here have presented,
will undoubtedly be extremely useful when developing and testing the next generation of PN simulations of BBHs in GCs.

\acknowledgments
It is a pleasure to thank M. Giersz, A. Askar, and B. Kocsis for
insightful discussions and comments on the manuscript.
J.S. acknowledges support from the Lyman Spitzer Fellowship.
D.J.D. acknowledges financial
support from NASA through Einstein Postdoctoral Fellowship award
number PF6-170151.

\bibliographystyle{h-physrev}
\bibliography{NbodyTides_papers}

\end{document}